%% file: body.tex
\documentclass{aa}
\usepackage{graphicx,psfig,amssymb,natbib,multirow}

\usepackage{txfonts}

\include{def}

\begin{document}

\title{The galaxy luminosity function and its evolution with \chandra} 

\author{P.~Tzanavaris \and  I.~Georgantopoulos}

\offprints{P.~Tzanavaris; email: pana@astro.noa.gr}

\institute{Institute of Astronomy \& Astrophysics, \ National
Observatory of Athens, I.~Metaxa \& V.~Pavlou, Penteli 152 36, Greece}

\date{Received... ;
accepted... }  

\abstract
{}
{We have compiled one of the largest normal-galaxy samples ever
to probe \x\ luminosity function evolution separately
for early and late-type systems.}
{We selected 207 normal galaxies up to redshift $z\sim 1.4$,
with data from four major \chandra\ \x\ surveys,
namely the \chandra\ deep fields (north, south and extended) and
XBootes,
and a combination of \x\ and optical criteria. 
We used template spectral energy-distribution fitting to obtain
separate early- and late-type sub-samples,
made up of 101 and 106 systems, respectively. 

For the full sample, as well as the
two sub-samples, we obtained
luminosity functions using both a non-parametric and
a parametric, maximum-likelihood method.}
{For the full sample, the non-parametric method 
strongly suggests luminosity evolution with redshift.
The maximum-likelihood estimate shows that this evolution
follows $\sim (1+z)^{k_{\rm total}}$, 
$k_{\rm total}=2.2\pm0.3$.
For the late-type sub-sample, we obtained 
$k_{\rm late}=2.4^{+1.0}_{-2.0}$.
We detected no significant evolution in the early-type
sub-sample. The distributions of early and late-type
systems with redshift show that late types dominate at
$z\ga 0.5$ and hence drive the observed evolution
for the total sample. 
}
{Our results 
support previous results in \x\ and other
wavebands, which suggests luminosity evolution with
$k=2-3$. 
}


\titlerunning{Galaxy luminosity function with \chandra}
\authorrunning{P.~Tzanavaris et al.}  

\maketitle

\section{Introduction}
Non-AGN dominated (\lq\lq normal\rq\rq) galaxies have intrinsically
weak \x\ luminosities and correspondingly faint fluxes. Thus, it is
only with the most recent, new-generation X-ray missions, \chandra\
and \xmm, that such galaxies have been detected in cosmologically
significant redshifts ($z>0$). The first such \x\ selected sample was
obtained with the \chandra\ Deep Fields North and South \citep[CDF-N,
CDF-S;][]{2003AJ....126..539A,2002ApJS..139..369G}, reaching fluxes of
$f(0.5-2.0 \ {\rm keV}) \sim 10^{-17}$~\funits \citep[see
also][]{2003AJ....126..575H}.  Based on these observations,
\citet{2004ApJ...607..721N} obtained an \x\ luminosity function
(XLF) of normal galaxies, characterised by luminosity evolution that
follows
$\sim (1+z)^{2.7}$. 

A population of rapidly evolving, star-forming galaxies has been
detected in recent years in other wavebands \citep[e.g.][and
references therein]{2004ApJ...615..209H}.  However, observations in
the \x\ band provide unique insight into physical processes,
complementing information obtained from other wavelength regions.  In
the most massive and luminous early-type galaxies, the X-ray emission
is dominated by the hot interstellar medium (ISM) at $kT\sim 1$~keV, with a
smaller fraction contributed by low-mass X-ray binaries (LMXBs)
associated with the old stellar population. However, 
for fainter \x\ systems, the evidence suggests that LMXBs
may well be the dominant X-ray emitting component
\citep[e.g.][and references therein]{2000ApJ...544L.101S,2003ApJ...586..826K}.

On the other hand, the X-ray
emission in late-type galaxies is mainly due to 
a mixture of low- and high-mass X-ray binaries
(HMXBs) with a contribution from hot gas ($kT\sim 1$~keV)
\citep[see][for a comprehensive review]{2006ARA&A..44..323F}.

Understanding how these populations of
binary stars and, consequently, their \x\ luminosity evolve with time
is obviously closely linked to XLF normal-galaxy evolution.
Considering LMXBs and HMXBs as the overall dominant component in \x\
emission from normal galaxies, \citet{2001ApJ...559L..97G} adopt a
semi-empirical approach to link \x-binary lifetimes with
star-formation rates (SFRs) in a cosmological context.  They show that
evolving SFRs significantly affect \x\ binary
populations, hence, integrated \x\ galactic emission, with the
possibility of significant evolution in \x\ luminosities even in
relatively low redshifts $\approxlt 1$ 
\citep[see also][]{1998ApJ...504L..31W}. They predict different
galaxy evolution rates at \x\ wavelengths, as compared to other
wavelength regions, which depend on the star-formation history of the
Universe, as well as evolutionary timescales for LMXBs and HMXBs. Conversely,
if the \x\ evolution of normal galaxies is known, one may obtain
insight into properties such as the characteristic timescales of LMXBs
and HMXBs. For instance, if HMXBs tracing the instantaneous SFR
dominate the integrated \x\ emission, the total \x\ luminosity is
expected to follow the star-formation history of the Universe as
observed in the optical and IR bands. On the contrary, LMXBs have much
longer evolutionary timescales, and the integrated \x\ luminosity of
normal galaxies would present a time delay of the order of $\sim
1$~Gyr compared to optical and IR observations. Constraining the
number density and evolution of LMXBs is also relevant to the LISA
gravitational wave mission, as this will be sensitive to gravitational
radiation from binaries with periods shorter than 4 hours. Such
sources are primarily LMXBs. Determining LMXB evolutionary timescales
can thus provide information on the number of expected LISA
detections.

As the relative contribution of LMXBs and HMXBs 
to the integrated \x\ luminosity of normal galaxies closely depends on
galaxy type, it is imperative to disentangle XLF behaviour
among different galaxy types. Unfortunately, because of the scarcity
of \x-detected normal galaxies, it is not surprising that very few
results have been obtained.  Using moderate size samples,
\citet{2005MNRAS.360..782G} and \citet{2006MNRAS.367.1017G} calculated
XLFs separately for early-type/absorption-line and
late-type/emission-line galaxies. By comparing the predictions of the
latter XLF with observed normal-galaxy number counts,
\citet{2006ApJ...641L.101G} detected luminosity evolution proportional
to $\sim (1+z)^{2.7}$ for {\it late} types, driven by sources with
\lxo~$>-2$. \citet{2007MNRAS.377..203G} then compared the
predictions of this XLF with observed counts for galaxies selected by
exploitation of the tight \x-IR correlation and detected luminosity
evolution proportional to $\sim (1+z)^{2.4}$ for their sample of
star-forming galaxies. Recently, \citet{arXiv:0706.1791v1} have used
the GOODS survey to obtain 40 early-type and
46 late-type galaxies up to a redshift $z\sim 1.2$.
Their XLFs suggest luminosity evolution proportional
to $\sim (1+z)^{1.6}$ for early types and 
$\sim (1+z)^{2.3}$ for late types.

In this paper we aim to substantially increase the numbers of \x\ detected
normal galaxies, to be able to investigate XLF evolution separately
for different galaxy types. We compiled one of the largest normal
galaxy samples ever, and, for the first time, probed redshift evolution
directly and independently for early and late-type systems.

The structure of this paper is as follows: In \scr{sec:sel} we
describe galaxy selection for our sample.  In \scr{sec:lf} we present
our XLFs, obtained using two different methods. 
We present our results in \scr{sec:res}, and discuss them
in \scr{sec:disc}. We conclude with predictions
related to future observations and missions in \scr{sec:fut}.

\section{Sample selection}\label{sec:sel}
We compiled our galaxy sample by 
cross-correlating four major X-ray surveys
with optical surveys overlapping in sky coverage.
We used  
the Extended \chandra\ Deep Field South
(E-CDF-S), the \chandra\ Deep Fields, North and South (CDF-N, CDF-S),
as well as the XBootes survey. We first identified X-ray and optical
counterparts and then used a number of selection criteria to
sift out normal-galaxy candidates from AGN.

\subsection{Galaxy selection criteria}
To apply the selection criteria described below, we required \x\
luminosities in the soft band 0.5\neg2.0 keV, \lx(0.5\neg2.0), as well
as hardness ratios, defined by
\begin{equation}\label{equ:HR}
{\rm HR}\equiv \frac{\rm H-S}{\rm H+S} \ .
\end{equation}
Here, $S$ represents counts in the soft energy
band, $0.5-2.0$~keV, and $H$ counts in the hard
energy band, $2.0-8.0$~keV.
We calculated the former by using the flux
information in the \x\ catalogues and the redshift information in the
optical catalogues. For E-CDF-S and CDF-N sources, we calculated
hardness ratios by using the soft, 0.5\neg2.0 keV, and hard,
2.0\neg8.0 keV, count information in the catalogues. For CDF-S and
XBootes, we used the harndess ratios given in the catalogues.

To separate NGs from AGN, we used all of the following criteria in conjunction:
\begin{itemize}
\item We demanded that X-ray sources be detected in
the 0.5-2.0 keV X-ray band, as non-AGN are preferentially soft
X-ray emitters \citep[e.g.,][]{2001ApJ...550..230L}. 
\item We imposed upper limits to 
\begin{itemize}
\item the logarithmic X-ray-to-optical flux
ratio, so that \logxr1. 
The optical flux $f_R$ was calculated separately
using the filter function for filters used in each survey.
This choice was motivated by
the fact that
some massive ellipticals will be missed by
a choice like \lxo~$\le-2$ \citep{2006A&A...454..447T}, combined with the
observation by \citet{2006ApJ...641L.101G} that
it is the \lxo~$>-2$ galaxies that drive XLF
evolution. The choice also took the
increasing importance of $k$-correcting with redshift
into account.

\item the X-ray luminosity, so that
$L_X < 10^{42}$~\lunits;
\item the hardness ratio, so that
HR~$\le 0$ (E-CDF-S and CDF-N), 0.0055 (CDF-S), or $-0.0078$ (XBootes). 
The last two values are due to the slightly different upper bounds of
the hard band in these catalogues and correspond to HR=0 for a 2.0\neg8.0
keV hard band and a power law $E^{-\Gamma}$ with $\Gamma=1.4$;
\end{itemize}
\item Finally, we used galaxy-type classification results from
the optical surveys, and performed individual visual checks
to keep sources that are clearly not stellar-like.

\end{itemize}

\subsection{X-ray and optical counterparts}

\subsubsection{E-CDF-S}
We cross-correlated the E-CDF-S \citep{2005ApJS..161...21L} with the
COMBO-17 survey \citep{2004A&A...421..913W}, identifying X-ray/optical
counterpart pairs the members of which are within 3\arcsec\ of each other.
By applying the galaxy selection criteria described above, we
identified 41 sources, with a median redshift \zmed~=~0.264.

\subsubsection{CDF-N}
We used counterparts identified by \citet{2003AJ....126..632B}, who
present optical and infra-red observations \citep[see
also][]{2004AJ....127..180C} of counterparts to X-ray sources of the
\chandra\ 2 Ms point-source catalogue \citep{2003AJ....126..539A}.
We thus selected 82 sources, with a median redshift \zmed~=~0.472.

\subsubsection{CDF-S}
The 1 Ms CDF-S catalogue is presented by \citet{2002ApJS..139..369G}.
We used spectroscopic and photometric information for counterparts
identified in \citet{2004ApJS..155..271S} and
\citet{2004ApJS..155...73Z}.  We selected 56 sources, with \zmed~=~0.52.

\subsubsection{XBootes}
We cross-correlated the XBootes X-ray point source catalogue
\citep{2005ApJS..161....9K} with the Sloan Digital Sky Survey, Data
Release 5 (SDSS, DR5), identifying \x/optical counterparts within
3\arcsec\ of each other. We thus identified 28 sources, with
\zmed~=~0.128.

In total, we obtained 207 sources up to $z\sim 1.4$. 
The total area curve for our sample is shown in \fr{fig:acurve}.
Figure~\ref{fig:lz} shows that 
the combination of deep and narrow surveys has the advantage of
providing wide coverage of the $L-z$ plane. 

\section{The luminosity function}\label{sec:lf}
\subsection{Non-parametric method}
We used the method of \citet{2000MNRAS.311..433P} to derive
the binned normal galaxy XLF. This is a variant of the classical
non-parametric $1/V_{\rm max}$ method \citep{1968ApJ...151..393S}
and has the advantage of being least affected by
systematic errors for sources close to the flux limit of the survey.

We estimated the function in a luminosity-redshift interval using the relation
\begin{equation}
\phi(L) = \frac{N}{\int_{L_{\rm min}} ^{L_{\rm max}}
                     \int_{z_{\rm min}(L)} ^{z_{\rm max}(L)}
                     \frac{dV}{dz} \, dz \, dL},
\end{equation}
where $N$ represents the number of sources with luminosities in the range
${L_{\rm min}}$ to ${L_{\rm max}}$, and $dV/dz$ is the
volume element for a redshift increment $dz$.
For a given luminosity $L$, $z_{\rm min}(L)$ and $z_{\rm max}(L)$
are the minimum and maximum redshifts for a source of that
luminosity to remain both within the flux limits of the survey and 
the redshift interval. Note that the solid angle $\Omega(L,z)$,
available for a source with luminosity $L$ at redshift $z$, corresponding
to a given flux in the area curve, also enters the calculation
via the volume element. The logarithmic bin size of
the function varies so that each bin comprises approximately equal
numbers of sources $N$.
From Poisson statistics, the uncertainty of
each luminosity bin is
\begin{equation}
\delta\phi(L,z) = \frac{\sqrt{N}}{\int_{L_{\rm min}} ^{L_{\rm max}}
                     \int_{z_{\rm min}(L)} ^{z_{\rm max}(L)}
                     \frac{dV}{dz} \, dz \, dL}.
\end{equation}
  
We first applied this method to our total sample. To investigate
luminosity function evolution, we estimated
the luminosity function independently in three
different redshift intervals, given in \tr{tab:z} together
with their corresponding median redshifts.
These were chosen empirically to
contain
roughly equal numbers of galaxies and
to bring out any evolutionary effects
as clearly as possible. 
The results are shown in 
\fr{fig:xlf-total.eps}
where there is a clear hint of evolution with redshift. We also note
redshift-dependent incompleteness because
less luminous sources are missed at higher redshift. 

As explained, we were interested in
investigating XLF behaviour as
a function of not only redshift, but also of galaxy
type. We separated our galaxies into early and late-type systems by
using broad-band colour information in several filters, obtained
from the optical catalogues.
We used the software {\it
hyperz} \citep{2000A&A...363..476B}, 
which performs \cs\ minimisation to select 
a template spectral energy-distribution (SED), 
which provides the best fit to a source's photometric SED. 
Although the primary goal of
{\it hyperz} is to obtain redshift information, the
best-fit template SED for a given source is equivalent 
to an early/late-type classification.
We used 61 template SEDs, which were smoothly
interpolated from four original galaxy SEDs, as described in
\citet{2004ApJS..155....1S}. These templates have indices that
increase with galaxy type from early to late. We classified
galaxies as early types, if the best-fit SED had index between 0 and
25, and as late types otherwise. We thus found that
our total sample was split into two roughly equal sub-samples,
comprising
101 early-type and 106 late-type systems. We then proceeded to estimate
the binned non-parametric luminosity function as before for
each sub-sample. As the number of galaxies was smaller by
a factor of $\sim 2$, 
we only used two redshift bins (\tr{tab:z}).

\subsection{Parametric method}
We also derived the luminosity function by means of the
parametric maximum-likelihood method \citep[ML;][]{1979ApJ...234..775T}.
The advantage of this method is its independence from 
a sample's homogeneity. It also allows us to quantify
the observed evolution by means of an evolution index, $k$, as
explained below. Its disadvantage is that the function
is assumed to have a certain form, with no possibility
of checking the goodness of fit. 
We adopted a \citet{1976ApJ...203..297S} form for the luminosity function,
\begin{equation}
\phi(L)dL = \phi^{\ast} \left(\frac{L}{L^{\ast}}\right)^{\alpha}
            \exp \left(-\frac{L}{L^{\ast}}\right) d\left(\frac{L}{L^{\ast}}
            \right) \, ,
\end{equation}
which is known to fit optical luminosity functions well. We parametrised
the characteristic luminosity $L^{\ast}$ where the function
changes from a power law with slope $\alpha$ to an exponential drop
at high luminosities as $L^{\ast}\equiv L^{\ast}_0 (1+z)^k$, where
the evolution index $k=0$ if $L^{\ast}=L^{\ast}_0$ at all redshifts.
The probability that a galaxy is detected with luminosity $L$ is
given by $P_i = \phi(L) /\int_{L_{\rm min}(z)}^\infty \phi(L^{\prime}) 
              dL^{\prime}$. We constructed the likelihood function
as $\prod_i P_i$ and maximised $\sum_i \ln P_i$ in $\log L$ space, by
varying $\alpha$, $L^{\ast}_0$ and $k$. Errors (90\%) were estimated 
from the regions about the ML fit where the likelihood changes by
1.3 \citep{1976ApJ...210..642A}. As the normalisation $\phi^{\ast}$
cancels out in this calculation, it was derived via
\begin{equation}
\phi^{\ast}= \frac{N}{\int\int \frac{\phi(L)}{\phi^{\ast}} 
             \frac{dV}{dz} dz\, dL} \, .
\end{equation}
Here, $N$ is the number of galaxies in the redshift interval, where ML
estimation had been performed.

To illustrate our ML fit results, we used the best-fit parameters to
calculate the luminosity function for the median redshifts that
correspond to the distinct redshift intervals used in the
{\it non}-parametric method, as explained in the next section.


\section{Results}\label{sec:res}
We plot the results of the non-parametric method in
Figures~\ref{fig:xlf-total.eps}, \ref{fig:xlf-early.eps},
\ref{fig:xlf-late.eps}, 
\ref{fig:evol0-04-early-late.eps}, and \ref{fig:evol04-14-early-late.eps}.
In the same plots we also show curves
illustrating the results of the ML fits.  These curves are produced
using the ML results shown in \tr{tab:mle} and the median redshifts of
each redshift interval used in the non-parametric
method (\tr{tab:z}).  
Note that Figs.~\ref{fig:xlf-early.eps} and
\ref{fig:xlf-late.eps} show the same results as
Figs.~\ref{fig:evol0-04-early-late.eps} 
and \ref{fig:evol04-14-early-late.eps}, the
difference being that the first pair is for different
galaxy types, whilst the second pair is for different redshift intervals.

The results of the
non-parametric method for the full sample (\fr{fig:xlf-total.eps})
show an indication of evolution with redshift. Although some
incompleteness inevitably sets in at higher redshift, the trend is
unmistakable. ML fitting quantifies this evolution with an index
$k_{\rm total}=2.2\pm0.3$.  The results for the early sample
(\fr{fig:xlf-early.eps}), however, tell a different story. First,
taking the errors into account, the results of the non-parametric
method offer no hint of evolution between
the two redshift intervals. This agrees with the results from
ML fitting, according to which the evolution index 
$k_{\rm early}=-0.7^{+1.4}_{-1.6}$ is
consistent with zero. The late-type results
(\fr{fig:xlf-late.eps}) complete the picture. 
There is an apparent gap between dark grey/red and light grey/green
points, which are obtained with the non-parametric method for different
redshift intervals.
According to the ML method, this
corresponds to an evolution index $k_{\rm late}=2.4^{+1.0}_{-2.0}$.

\section{Discussion}\label{sec:disc}
\subsection{Comparison with previous results}
To probe XLF evolution for different galaxy types, large numbers are
necessary. \citet{2005MNRAS.360..782G} used 46 galaxies to
construct XLFs for emission- and absorption-line
systems. However, their results were for a single redshift bin
$z<0.22$. \citet{2006MNRAS.367.1017G} had 67 galaxies, likewise limited
to a single redshift bin $z<0.2$. Although \citet{2004ApJ...607..721N}
had a large sample of 210 galaxies, they constructed no
galaxy-type specific sub-samples. It is thus the first time that
XLFs were calculated directly for early and late-type normal galaxies.
The large size of our total sample allows splitting into three
redshift bins that, in turn, strongly suggest evolution with
redshift. After 
repeating the procedure separately for early and late-type systems,
we detected no evolution for early types and strong evolution
for late types. 

Our results are mostly in good agreement with related work in the
literature. The results for 36 normal galaxies ($z=0.01
\rightarrow 0.3$) by \citet{2006ApJ...644..829K} are shown by stars
in \fr{fig:xlf-total.eps}. The results from
\citet{2004ApJ...607..721N} for two redshift bins are also
shown. When compared to our results, these show an apparent systematic
shift to higher luminosities, which might suggest AGN
contamination. Note, however, that for the sake of 
clarity we do not show errors
from \citet{2004ApJ...607..721N}, which would bring their points and, in
particular, the apparently markedly discrepant point at $(\log L_X, \,
\log \, \phi) \sim (40.2,-1.7)$ into better agreement with ours.

In Figs.~\ref{fig:xlf-early.eps} and \ref{fig:xlf-late.eps} we also
show the results for early- and late-type galaxies by
\citet{2006MNRAS.367.1017G}. We constructed the XLF curves by using the
function parameters quoted by these authors, but also imposed
luminosity evolution following the $k$ values we found in the present
paper. For early types we see that our largest sample cannot be
adequately parametrised by their XLF, which shows too steep an
exponential cutoff. 
The late-type XLF shows broad
agreement with our non-parametric bins, but it is clearly not a good
representation of our data. The discrepancies, especially for
early types, may partly stem from these
authors using a selection criterion
\lxol2, thus obtaining fewer luminous sources.


\subsection{Evolution}
The fact that late-type galaxies are driving the observed evolution of
the total sample can be understood if we look at the redshift
distributions of the two galaxy types.  The histograms in
\fr{fig:Nz.eps} show the observed distributions: Early-types dominate
in the lowest redshift bins, whilst late-types dominate at
intermediate redshifts. A Kolmogorov-Smirnov (KS) test shows that
the two distributions differ significantly.
The probability that the value of the KS statistic $D$
obtained may be due to chance alone is very small, $p=0.006$.  The
observed trend is also corroborated by theoretical
distributions, shown by curves in the same figure, which we construct
using our ML fit parameters together with the area curve information
of our data. Qualitatively, this also agrees with
the results of \citet{2005ApJ...625..621B}. These authors
used a large sample from the Great Observatories Origins
Deep Survey (GOODS) fields to probe mass assembly of
morphologically distinct normal galaxies. They found
an increasing proportion of early-type systems with
decreasing redshift from $z\sim 1$ to 0, which is
similar to what we see in our redshift distribution.

Furthermore, we found that it is the late types with highest \lxo\ that
are mostly responsible for the observed trend. In \fr{fig:Nz.eps} we
see that numbers for late-types appear to rise after $z=0.5$. Out of
53 late-type systems at $z>0.5$, 39, or 74\%, have $\log (f_X/f_R) >
-2$. This finding agrees with the claim by
\citet{2006MNRAS.367.1017G} that their deduced evolution stems
primarily from such sources.  It also provides {\it a posteriori}
justification of our upper-limit choice ($\log (f_X/f_R) \le -1$)
in the selection criteria for
the present sample.  We cannot exclude that this observation is, at
least partially, a consequence of a selection effect.
Given the steep relation $L_X\propto L_B^{1.5}$ 
for spirals \citep{2001ApJS..137..139S}, we might expect to
preferentially detect intrinsically brighter galaxies at higher
redshift, which would lead to a selection against low $f_X/f_R$ values.
Even so, that we see bright {\it late}-types may
still be significant.
As \citet{2006A&A...454..447T} have shown, at lower redshift,
bright galaxies with \lxo~$>-2$ are mostly early-type.

It is encouraging that our luminosity evolution estimates are in good
agreement with results by other authors.  Taken together,
\citet{2004ApJ...607..721N}, \citet{2004ApJ...615..209H},
\citet{2005A&A...440...23R}, \citet{2006ApJ...641L.101G},
\citet{2007MNRAS.377..203G}, \citet{arXiv:0706.1791v1},
as well as the present paper, lead to a
broad consensus that {\it the data are consistent with luminosity
evolution with an index $k=2-3$.}

In particular, \citet{2004ApJ...615..209H} investigates
luminosity function evolution for
a sample of radio-selected star-forming galaxies by combining
constraints from the global SFR density
evolution with those from the 1.4 GHz radio source counts 
at submillijansky levels.
He finds luminosity evolution $\sim (1+z)^{2.7\pm0.6}$ for
star-forming galaxies. Given the tight correlation
between radio, far infrared, and \x\ luminosity functions
\citep{2003A&A...399...39R}, the agreement of this result
with our result for late-type systems is particularly 
significant for two reasons. Firstly, it supports our 
main result that
late-type systems are driving the overall redshift
evolution of normal galaxies in the local Universe. Secondly,
as our normal galaxy selection criteria are independent of
those for 1.4 GHz sources, it suggests that any AGN contamination
is unlikely to have had significant impact on our XLFs.

\citet{2007MNRAS.377..203G}
obtained essentially the same evolution index for late-types, $k_{\rm
late}=2.4$, by employing methods largely independent of
ours.
Because they used \x-IR correlations which hold for
normal galaxies, but not for AGN, their sample is virtually guaranteed to
be free of AGN contamination. Although their sample is $\sim 4$ times
smaller than ours, the agreement in the evolution
index is, once more, significant.
\citet{2005A&A...440...23R} showed that the blue
galaxy luminosity function for the 25,000-strong
spiral galaxy sample of \citet{2003A&A...401...73W} 
requires a pure luminosity evolution index
$\la 3$. This is also consistent with our results,
under the assumption that blue galaxies 
are largely the same population as the late-type
systems in our sample.
\citet{arXiv:0706.1791v1} jointly fitted low
and high-redshift XLFs, obtaining
pure-luminosity evolution with 
$k_{\rm early} = 1.6$ and 
$k_{\rm late} = 2.3$. These authors used
Bayesian techniques and a Monte Carlo Markov chain
analysis. The agreement, at least for late-types,
is thus particularly encouraging.

Comparisons with results for red galaxies are less straightforward.
Assuming that our early-type galaxies are the same population as
red galaxies in optical surveys, with no significant dust
contamination, 
that, unlike \citet{arXiv:0706.1791v1}, we do not measure
significant evolution for early-type galaxies, disagrees 
at face value 
with the observed evolution in optical
wavebands
\citep{2007ApJ...654..858B,2003A&A...401...73W}. However, this is 
consistent with models that predict a fast
decline in luminosities at optical wavelengths, with
\x\ luminosities remaining high for $\ga 1$ Gyr
\citep{2006IAUS..230..417E}, in systems with a significant
contribution from LMXBs. 
\citet{1998ApJ...504L..31W} predict that the
combined effect of LMXBs and HMXBs may lead
to a \lq twin peak\rq\ in the evolving XLF
of normal galaxies; due to
HMXBs, the first peak would be expected to occur close
to the SFR peak at $z\approx 1.5$. The
second peak would be 
delayed until $z\approx 0.5-1$,
due to delayed turn-on of
LMXBs.
To first order,
our early-type systems are dominated by LMXBs, which
could mean that we
may approximately be witnessing this second peak.
However, this approximation may be inadequate, even
to first order, due to the caveat 
that the most luminous systems
are dominated by the hot ISM.
Similarly, our late-type systems are
dominated by HMXBs, whose \x\ luminosity
has peaked at high $z$, and we are witnessing
their dimming at lower redshifts. Indeed, an
evolution index $k\sim 2.4$ is consistent with
\lq Peak-M\rq\ models, which are characterised
by longer evolutionary timescales for LMXBs
\citep[][Table 2]{2001ApJ...559L..97G}.

\subsection{Comparison with AGN}
It is not clear whether this scenario is
consistent with results for the XLF evolution
of AGN-dominated systems. 
\citet{2003ApJ...598..886U} and
\citet{2005ApJ...635..864L} found
that luminosity-dependent density evolution
explains their XLFs best.
They showed that XLFs for luminous AGN peak
at higher redshifts than for less luminous ones.
By assuming that SMBH growth is closely
linked to starburst activity,
\citet{2003ApJ...598..886U} linked this to
normal-galaxy evolution in a first-order
scenario, where luminous AGN once lived in what
later became early-type galaxies. Their
AGN activity peaked at high $z\approx 2$, 
following strong starbursts,
and decreased
rapidly after $z\la 2$. On the other hand,
galaxies that hosted less luminous AGN at high $z$
have smaller spheroids, i.e. are of later type. 
Their star-formation peaks at lower $z$, and so
does their AGN activity, peaking at $z\approx 0.6-0.7$. 

We did not specifically investigate luminosity-dependent
density
evolution in this paper. As explained in the
previous section, our results supporting pure-luminosity
evolution are corroborated by a number of different authors
using a variety of data and techniques. This would then
cast doubt on a general, unified evolutionary scheme,
encompassing both AGN and normal galaxies.

Even so, it is possible to make our results broadly consistent
with these AGN evolutionary models, at least in a
qualitative sense. Thus early types that, presumably, 
correspond to bright AGN/high star-formation at high redshift 
would be expected to show declining luminosities at the
redshifts we probe. That we detect no such
evolution is already explained by the possibility
of delayed LMXB evolution. For late types, we
do detect declining luminosities. As the
AGN peak redshift is within the range probed by our
sample, our selection criteria naturally exclude
any AGN, instead picking galaxies whose AGN is either
switching off or makes a very low contribution to the
total luminosity.



%
%
%
%

\section{Predictions for deeper \chandra\
and {\it XEUS} observations}\label{sec:fut}
The CDFs have allowed us, for the first time,
to obtain a couple of hundred \x-detected 
normal galaxies at cosmologically
interesting redshifts. In turn, we are now
starting to detect galaxy-type
dependent evolution. However,
compared to results in other
(e.g. optical) wavelength bands, 
\x\ work still suffers severely from
small numbers. Incompleteness at
high redshift is evident, and
uncertainties large. It is thus imperative
to obtain larger and deeper samples for
exploring normal galaxy evolution in
greater detail. 

In \fr{fig:lognlogs.eps} we present results
from various number count distributions
as a function of soft \x\ flux including
observed CDF
counts, which come both from AGN and normal galaxies, 
observed galaxy counts, and
a theoretical $N(S)$
distribution, which we calculated using
our MLE results.
With
data from a 4 Ms CDF-N
exposure, $\sim$~\ten{6}{3} sources per
square degree would
be detected with fluxes $\ga$~\ten{1.2}{-17}~\funits.
Galaxies with luminosities log \lx~$\sim 41$ would
then be detectable out to $z\sim 1.4$. The
brightest galaxies, log \lx~$\sim 42$ would
extend the redshift range to $z\la 4$.

Such exposure-time requirements
may well prove prohibitive in practice.
ESA's planned X-ray Evolving Universe Spectroscopy
({\it XEUS}) mission will be able to offer an
impressive increase in detected number within
realistic time constraints. Using our evolving
luminosity function
with no evolutionary break at high redshift, we estimate that,
just with a single 1 Ms exposure, counts
from normal galaxies 
(see \fr{fig:lognlogs.eps})
will exceed
$\sim$~\ten{2}{4} per square degree for
sources with fluxes $\ga$~\ten{4}{-18}~\funits.
In this case log \lx~$\sim 41$ galaxies will be detectable to
$z\sim2$ and log \lx~$\sim 42$ ones to $z\sim 6$, thus
finally paving the way for a study of the high-redshift
normal-galaxy XLF.
%
This is consistent with the fluctuation analysis results
of \citet{2002ApJ...564L...5M}, who show that soft-band
counts for all sources (i.e. including AGN) continue growing with
decreasing flux. As our analysis specifically targets
normal galaxies, this result implies that counts from normal
galaxies will overtake AGN at fluxes
$\sim$~\ten{5}{-18}~\funits.


\section{Summary}  
In this paper we have shown the importance of increasing
numbers of \x-luminous normal galaxies for probing
XLF behaviour as a function of both redshift and
galaxy type. 
We selected 207 normal galaxies (101 early-type
and 106 late-type) from the three \chandra\ deep
fields and XBootes.
Our major results are:
\begin{enumerate}
\item Both methods of XLF estimation,
parametric and non-parametric
suggest that the full galaxy sample is
consistent with pure luminosity evolution, which is
driven exclusively by evolution of late-type systems. 
\item The evolution index $k$, where 
$L^{\ast}\equiv L^{\ast}_0 (1+z)^k$, is 
$2.2\pm0.3$ for the full sample, 
\aer{2.4}{+1.0}{-2.0} for the late-type sample, and
\aer{-0.7}{+1.4}{-1.6} for the early-type sample.
\item Our results agree broadly with results from other work.
\item We estimate that, with a 1 Ms exposure with {\it XEUS}, 
counts from normal galaxies will overtake AGN at faint fluxes and
will allow probing XLF evolution to high-redshift.
\end{enumerate}

\section{Acknowledgements} 
We thank Antonis Georgakakis for making his luminosity function
fitting code available to us.
This work is funded in part by the Greek National Secretariat for
Research and Technology within the framework of the Greece-USA
collaboration programme {\it Study of Galaxies with the Chandra X-ray
Satellite}.  We acknowledge the use of data from the \xmm\ Science
Archive at VILSPA, the \chandra-XAssist archive, and the \2df. This
research has made use of data obtained from the High Energy
Astrophysics Science Archive Research Center (HEASARC), provided by
NASA's Goddard Space Flight Center. This research made use of the
NASA/IPAC Extragalactic Database (NED) operated by the Jet
Propulsion Laboratory, California Institute of Technology, under
contract with the National Aeronautics and Space Administration.

Funding for the Sloan Digital Sky Survey (SDSS) and SDSS-II is
provided by the Alfred P. Sloan Foundation, the Participating
Institutions, the National Science Foundation, the U.S. Department
of Energy, the National Aeronautics and Space Administration, the
Japanese Monbukagakusho, the Max Planck Society, and the
Higher Education Funding Council for England. The SDSS Web site is
http://www.sdss.org/.
The SDSS is managed by the Astrophysical Research Consortium (ARC)
for the Participating Institutions, which are
are the American Museum of Natural History, Astrophysical
Institute Potsdam, University of Basel, University of Cambridge,
Case Western Reserve University, The University of Chicago, Drexel
University, Fermilab, the Institute for Advanced Study, the Japan
Participation Group, The Johns Hopkins University, the Joint
Institute for Nuclear Astrophysics, the Kavli Institute for
Particle Astrophysics and Cosmology, the Korean Scientist Group,
the Chinese Academy of Sciences (LAMOST), Los Alamos National
Laboratory, the Max-Planck-Institute for Astronomy (MPIA), the
Max-Planck-Institute for Astrophysics (MPA), New Mexico State
University, Ohio State University, University of Pittsburgh,
University of Portsmouth, Princeton University, the United States
Naval Observatory, and the University of Washington.

\tiny

\include{fig}

\include{tabl}
\end{document}

%% file: def.tex
\def\funits{erg~cm$^{-2}$~s$^{-1}$}

\def\lunits{erg~s$^{-1}$}

\def\cs{$\chi^2$}

\def\logxr1{$\log (f_X/f_R) \le -1$}

\def\lstar{$L^{\ast}$}
\def\lx{$L_X$}
\def\lxo{$\log (f_X/f_O)$}
\def\lxol2{$\log (f_X/f_O) < -2$}
\def\neg{$-$}

\def\phistar{$\phi^{\ast}$}
\def\ran{$-2 < \log (f_X/f_O) < -1$}
\def\ran1{$-1 < \log (f_X/f_O) < +1$}

\def\x{X-ray}

\def\zmed{$z_{\rm med}$}

\def\chandra{{\it Chandra}}

\def\2df{{\it 2dfGRS}}
\def\xmm{{\it XMM-Newton}}

\def\fr{Fig.~\ref}
\def\scr{Sect.~\ref}
\def\tr{Table~\ref}

\def\exi{\begin{equation}}
\def\exo{\end{equation}}

\newcommand{\aer}[3]{$#1^{#2}_{#3}$} 
\newcommand{\ten}[2]{$#1\times 10^{#2}$} 

\def\spose#1{\hbox to 0pt{#1\hss}} 
\def\approxlt{\mathrel{\spose{\lower 3pt\hbox{$\sim$}}
        \raise 2.0pt\hbox{$<$}}}
\def\approxgt{\mathrel{\spose{\lower 3pt\hbox{$\sim$}}
        \raise 2.0pt\hbox{$>$}}}


%% file: fig.tex
\begin{figure}
\includegraphics[width=9cm]{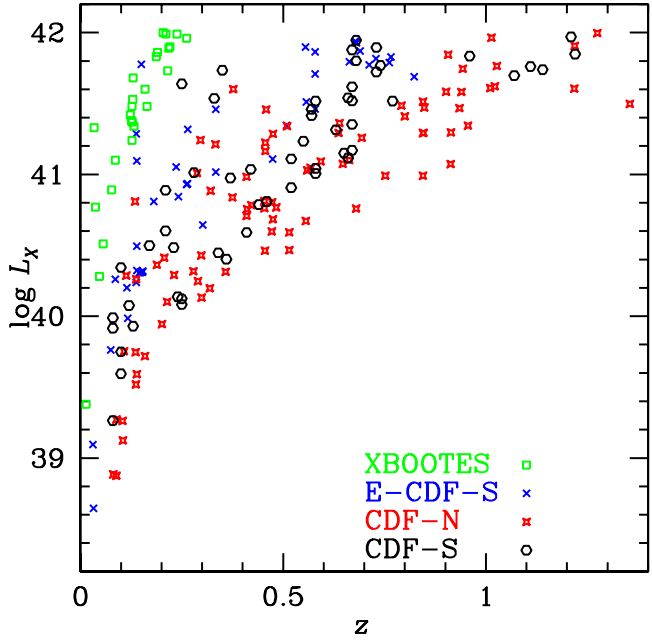}
\caption{Luminosity-redshift plot for the total sample used in this work. 
The positions of galaxies selected from the four different sub-samples 
are shown with
the different symbols and colours indicated.}
\label{fig:lz}
\end{figure}

\begin{figure}
\includegraphics[width=9cm]{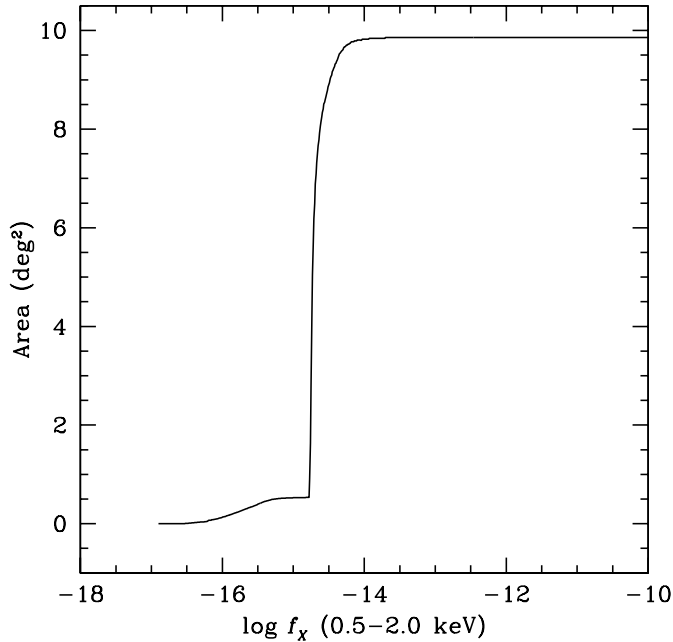}
\caption{Area covered as a function of flux limit for the total sample.}
\label{fig:acurve}
\end{figure}

\begin{figure}
\includegraphics[width=9cm]{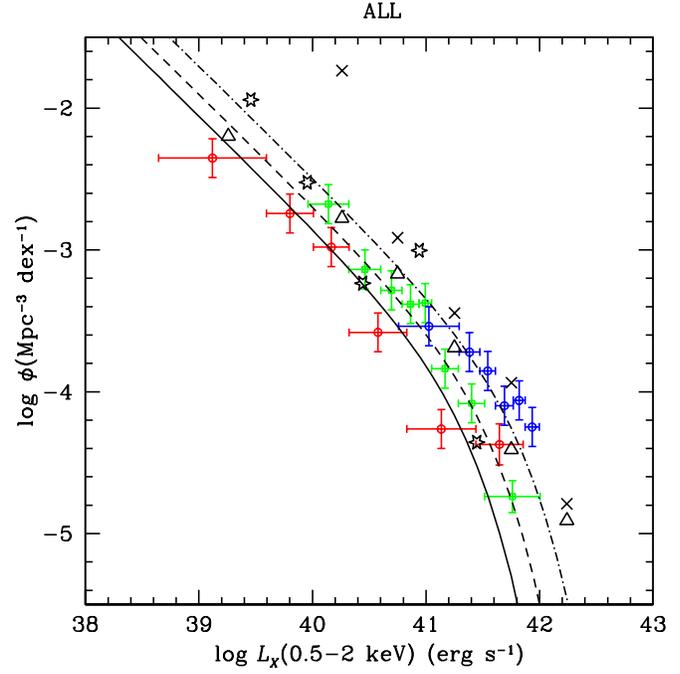}
\caption{Luminosity function for the total sample. 
{\it Non-parametric method} \citep{2000MNRAS.311..433P}:
Points with error bars indicate
$\log \phi$ values
estimated for galaxies selected in a low (red/medium-grey circles), 
intermediate (green/light-grey squares), and high (blue/dark-grey circles)
redshift interval. 
Horizontal error bars 
indicate the size
of the logarithmic luminosity bin used, and vertical error bars the Poisson
error.
{\it Parametric method (ML)}:
For the same redshift intervals,
the three curves (solid, dashed, dot-dashed, in increasing 
redshift order) show the luminosity function calculated from our ML
fit parameters shifted to the median redshift of
each interval (see text for details).
{\it Other results}:
Stars are from the work of 
\citet{2006ApJ...644..829K}.
The results of \citet{2004ApJ...607..721N} are shown with 
crosses ($z>0.5$)
and triangles ($z<0.5$).}
\label{fig:xlf-total.eps}
\end{figure}

\begin{figure}
\includegraphics[width=9cm]{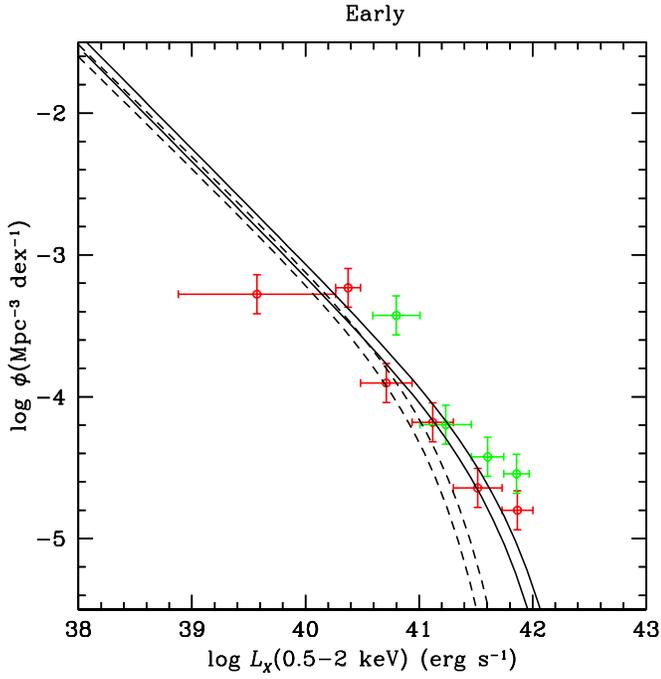}
\caption{Luminosity function for the sub-sample of early types.
{\it Points with error bars}:
results from the \citet{2000MNRAS.311..433P} method
in a low- (dark grey/red) and high-redshift (light grey/green)
interval, as explained in the text (see also \tr{tab:z}).
{\it Solid curves}: results from our ML fits shifted
to the median redshifts of the
same redshift intervals. {\it Dashed curves}: 
luminosity function from \citet{2006MNRAS.367.1017G} shifted
to the median redshifts as explained in the text.}
\label{fig:xlf-early.eps}
\end{figure}
 
\begin{figure}
\includegraphics[width=9cm]{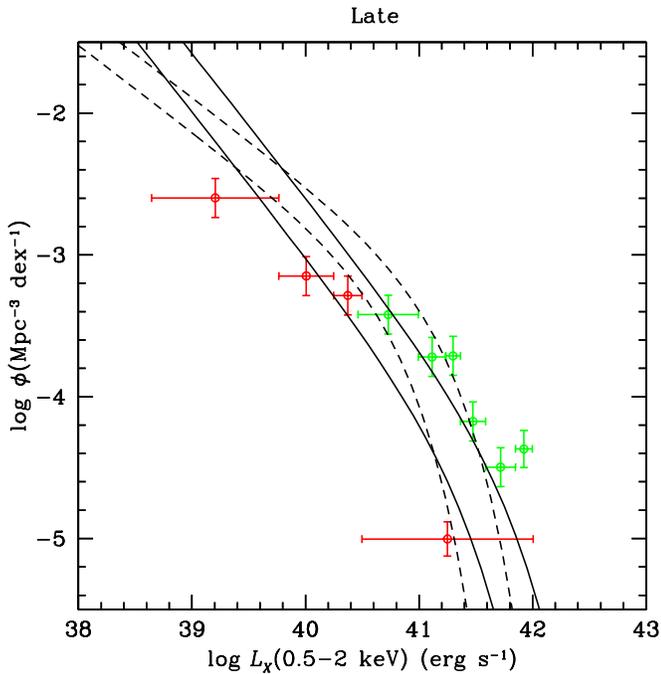}
\caption{Like \fr{fig:xlf-early.eps} but 
for the sub-sample of late types.}
\label{fig:xlf-late.eps}
\end{figure} 

\begin{figure}
\includegraphics[width=9cm]{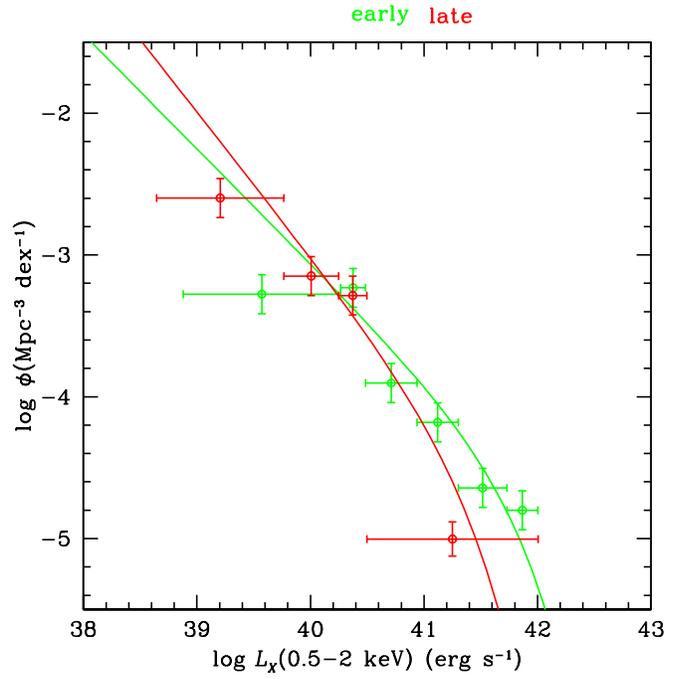}
\caption{Luminosity function in the redshift
interval $0-0.4$. 
{\it Points with error bars}: results
from the \citet{2000MNRAS.311..433P} method for
early (green/light grey) and late (red/dark grey) types.
{\it Solid curves}: results from ML fits shifted
to the median redshift of the redshift interval 
(same colour coding as for points).}
\label{fig:evol0-04-early-late.eps}
\end{figure}

\begin{figure}
\includegraphics[width=9cm]{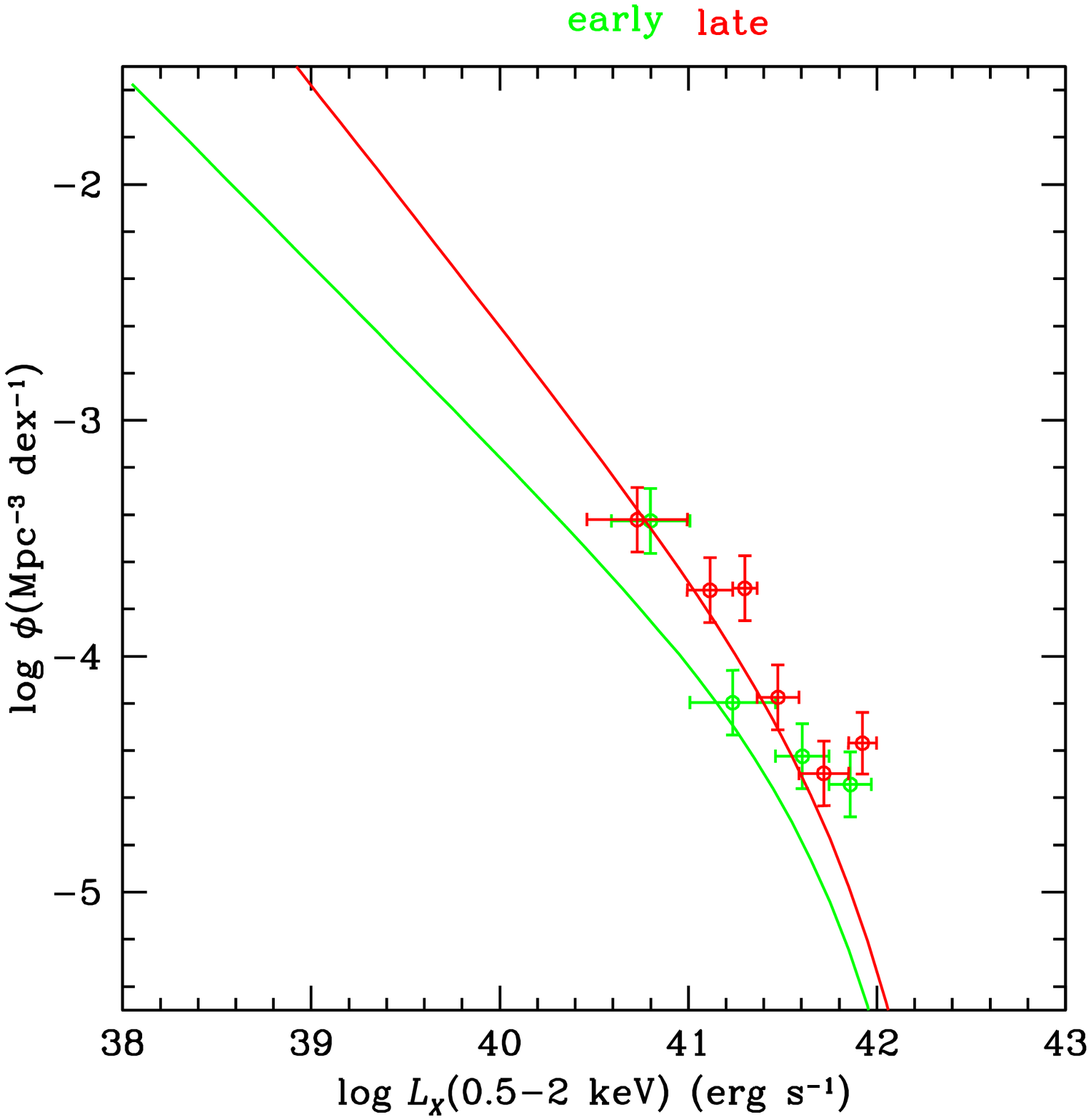}
\caption{Like \fr{fig:evol0-04-early-late.eps}, but
for the redshift interval $0.4-1.4$.}
\label{fig:evol04-14-early-late.eps}
\end{figure}

\begin{figure}
\includegraphics[width=9cm]{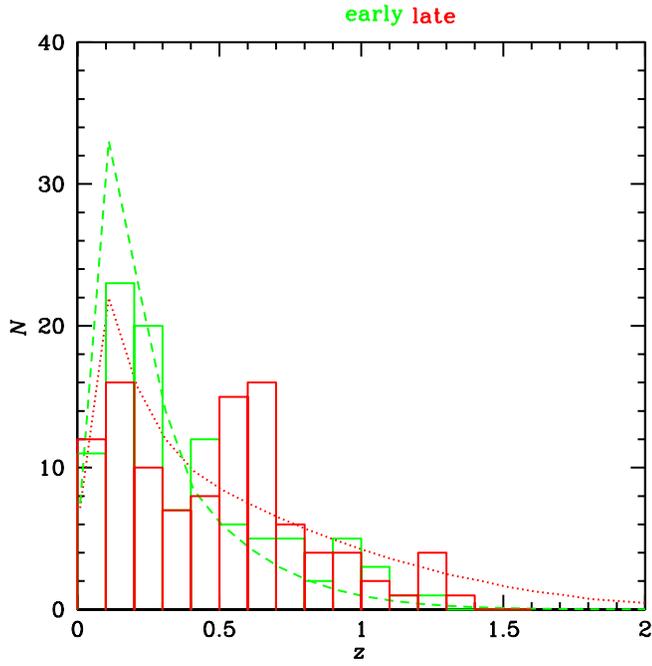}
\caption{Redshift distributions. {\it Histograms}:
distributions for the sub-samples of early (green/light grey) and
late-type (red/dark grey) galaxies. {\it Curves}:
calculated distributions for early (green/dashed) and late (dotted/red) types.}
\label{fig:Nz.eps}
\end{figure}

\begin{figure}
\includegraphics[width=9cm]{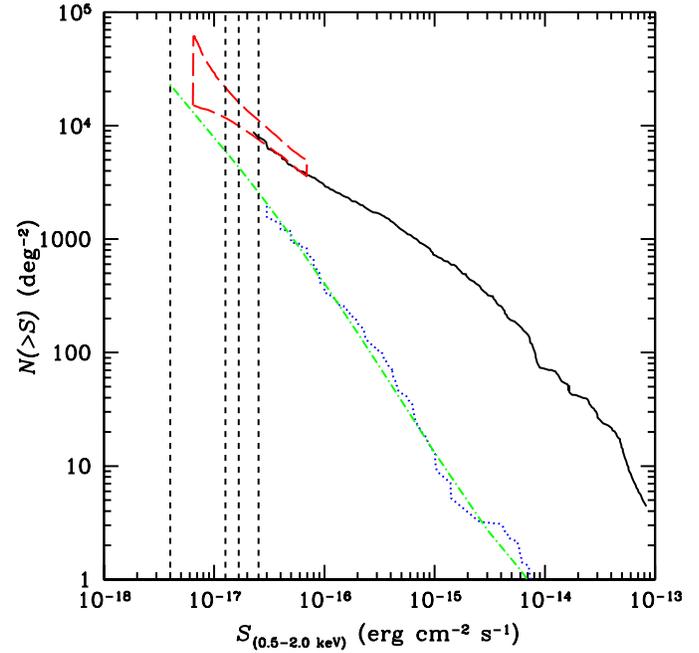}
\caption{Number of sources per square degree as a function of flux limit.
{\it Solid black curve}: total counts from CDFs 
\citep{2004AJ....128.2048B}. {\it Red long-dash bounded area}: 1 Ms
fluctuation analysis results \citep{2002ApJ...564L...5M}.
{Blue dotted curve}: normal galaxies from this work.
{Green dot-dashed curve}: counts calculated using our estimated
values for $L^{\ast}_0$, $\alpha$ and $k$ for the 
normal galaxy luminosity function.
From left to right, the vertical dashed lines show flux
limits for a 1 Ms exposure with {\it XEUS}, a 4 Ms CDF-N, a 3 Ms CDF-N,
and the 2 Ms CDF-N. }
\label{fig:lognlogs.eps}
\end{figure}

%% file: tabl.tex
\begin{table} 
\caption{Information on redshift intervals used for the parametric XLF estimation method.} 
\label{tab:z} 
\centering 
\begin{tabular}{|c|c|c|} 
\hline 
Sample & $z$ range & \zmed\ \\
\hline
\multirow{3}{*}{Total} & $0\rightarrow 0.2$   & 0.13 \\
                       & $0.2\rightarrow 0.6$ & 0.38   \\
                       & $0.6\rightarrow 1.4$ & 0.78  \\ \hline
\multirow{2}{*}{Early} & $0\rightarrow 0.4$   & 0.17  \\
                       & $0.4\rightarrow 1.4$ & 0.67   \\ \hline
\multirow{2}{*}{Late} & $0\rightarrow 0.4$   & 0.14 \\
                       & $0.4\rightarrow 1.4$ & 0.67   \\

\hline

\end{tabular} 
\end{table}
\begin{table}
\caption{ML fit results for a Schechter function with evolution index $k$.}

\label{tab:mle}
\begin{tabular}{ccc cc}
\hline
Sample & log \lstar\                 & $\alpha$           & \phistar\                   & $k$ \\ 
       & (\lunits)                   &        & $\ln (10) \times 10^{-4}$ Mpc$^{-3}$ dex$^{-1}$  &\\
\hline
Total  & \aer{41.24}{+0.02}{-0.02}   & \aer{-1.79}{+0.06}{-0.07}  & \aer{1.24}{}{}       & \aer{2.2}{+0.3}{-0.3}\\
Early  & \aer{41.87}{+0.4}{-0.3}     & \aer{-1.81}{+0.17}{-0.18}  & \aer{0.295}{}{}       & \aer{-0.7}{+1.4}{-1.6}\\
Late   & \aer{41.25}{+0.50}{-0.25}    & \aer{-2.02}{+0.12}{-0.09}  & \aer{0.379}{}{}     & \aer{2.4}{+1.0}{-2.0}\\ 
\hline
\end{tabular}
\end{table}

%% file: body.bbl
\begin{thebibliography}{43}
\expandafter\ifx\csname natexlab\endcsname\relax\def\natexlab#1{#1}\fi

\bibitem[{{Alexander} {et~al.}(2003){Alexander}, {Bauer}, {Brandt},
  {Schneider}, {Hornschemeier}, {Vignali}, {Barger}, {Broos}, {Cowie},
  {Garmire}, {Townsley}, {Bautz}, {Chartas}, \&
  {Sargent}}]{2003AJ....126..539A}
{Alexander}, D.~M., {Bauer}, F.~E., {Brandt}, W.~N., {et~al.} 2003, \aj, 126,
  539

\bibitem[{{Avni}(1976)}]{1976ApJ...210..642A}
{Avni}, Y. 1976, \apj, 210, 642

\bibitem[{{Barger} {et~al.}(2003){Barger}, {Cowie}, {Capak}, {Alexander},
  {Bauer}, {Fernandez}, {Brandt}, {Garmire}, \&
  {Hornschemeier}}]{2003AJ....126..632B}
{Barger}, A.~J., {Cowie}, L.~L., {Capak}, P., {et~al.} 2003, \aj, 126, 632

\bibitem[{{Bauer} {et~al.}(2004){Bauer}, {Alexander}, {Brandt}, {Schneider},
  {Treister}, {Hornschemeier}, \& {Garmire}}]{2004AJ....128.2048B}
{Bauer}, F.~E., {Alexander}, D.~M., {Brandt}, W.~N., {et~al.} 2004, \aj, 128,
  2048

\bibitem[{{Bolzonella} {et~al.}(2000){Bolzonella}, {Miralles}, \&
  {Pell{\'o}}}]{2000A&A...363..476B}
{Bolzonella}, M., {Miralles}, J.-M., \& {Pell{\'o}}, R. 2000, \aap, 363, 476

\bibitem[{{Brown} {et~al.}(2007){Brown}, {Dey}, {Jannuzi}, {Brand}, {Benson},
  {Brodwin}, {Croton}, \& {Eisenhardt}}]{2007ApJ...654..858B}
{Brown}, M.~J.~I., {Dey}, A., {Jannuzi}, B.~T., {et~al.} 2007, \apj, 654, 858

\bibitem[{{Bundy} {et~al.}(2005){Bundy}, {Ellis}, \&
  {Conselice}}]{2005ApJ...625..621B}
{Bundy}, K., {Ellis}, R.~S., \& {Conselice}, C.~J. 2005, \apj, 625, 621

\bibitem[{{Capak} {et~al.}(2004){Capak}, {Cowie}, {Hu}, {Barger}, {Dickinson},
  {Fernandez}, {Giavalisco}, {Komiyama}, {Kretchmer}, {McNally}, {Miyazaki},
  {Okamura}, \& {Stern}}]{2004AJ....127..180C}
{Capak}, P., {Cowie}, L.~L., {Hu}, E.~M., {et~al.} 2004, \aj, 127, 180

\bibitem[{{Eracleous} {et~al.}(2006){Eracleous}, {Sipior}, \&
  {Sigurdsson}}]{2006IAUS..230..417E}
{Eracleous}, M., {Sipior}, M.~S., \& {Sigurdsson}, S. 2006, in IAU Symposium,
  Vol. 230, Populations of High Energy Sources in Galaxies, ed. E.~J.~A.
  {Meurs} \& G.~{Fabbiano}, 417--422

\bibitem[{{Fabbiano}(2006)}]{2006ARA&A..44..323F}
{Fabbiano}, G. 2006, \araa, 44, 323

\bibitem[{{Georgakakis} {et~al.}(2006{\natexlab{a}}){Georgakakis},
  {Georgantopoulos}, {Akylas}, {Zezas}, \& {Tzanavaris}}]{2006ApJ...641L.101G}
{Georgakakis}, A., {Georgantopoulos}, I., {Akylas}, A., {Zezas}, A., \&
  {Tzanavaris}, P. 2006{\natexlab{a}}, \apjl, 641, L101

\bibitem[{{Georgakakis} {et~al.}(2007){Georgakakis}, {Rowan-Robinson},
  {Babbedge}, \& {Georgantopoulos}}]{2007MNRAS.377..203G}
{Georgakakis}, A., {Rowan-Robinson}, M., {Babbedge}, T.~S.~R., \&
  {Georgantopoulos}, I. 2007, \mnras, 377, 203

\bibitem[{{Georgakakis} {et~al.}(2006{\natexlab{b}}){Georgakakis},
  {Chavushyan}, {Plionis}, {Georgantopoulos}, {Koulouridis}, {Leonidaki}, \&
  {Mercado}}]{2006MNRAS.367.1017G}
{Georgakakis}, A.~E., {Chavushyan}, V., {Plionis}, M., {et~al.}
  2006{\natexlab{b}}, \mnras, 367, 1017

\bibitem[{{Georgantopoulos} {et~al.}(2005){Georgantopoulos}, {Georgakakis}, \&
  {Koulouridis}}]{2005MNRAS.360..782G}
{Georgantopoulos}, I., {Georgakakis}, A., \& {Koulouridis}, E. 2005, \mnras,
  360, 782

\bibitem[{{Ghosh} \& {White}(2001)}]{2001ApJ...559L..97G}
{Ghosh}, P. \& {White}, N.~E. 2001, \apjl, 559, L97

\bibitem[{{Giacconi} {et~al.}(2002){Giacconi}, {Zirm}, {Wang}, {Rosati},
  {Nonino}, {Tozzi}, {Gilli}, {Mainieri}, {Hasinger}, {Kewley}, {Bergeron},
  {Borgani}, {Gilmozzi}, {Grogin}, {Koekemoer}, {Schreier}, {Zheng}, \&
  {Norman}}]{2002ApJS..139..369G}
{Giacconi}, R., {Zirm}, A., {Wang}, J., {et~al.} 2002, \apjs, 139, 369

\bibitem[{{Hopkins}(2004)}]{2004ApJ...615..209H}
{Hopkins}, A.~M. 2004, \apj, 615, 209

\bibitem[{{Hornschemeier} {et~al.}(2003){Hornschemeier}, {Bauer}, {Alexander},
  {Brandt}, {Sargent}, {Bautz}, {Conselice}, {Garmire}, {Schneider}, \&
  {Wilson}}]{2003AJ....126..575H}
{Hornschemeier}, A.~E., {Bauer}, F.~E., {Alexander}, D.~M., {et~al.} 2003, \aj,
  126, 575

\bibitem[{{Kenter} {et~al.}(2005){Kenter}, {Murray}, {Forman}, {Jones},
  {Green}, {Kochanek}, {Vikhlinin}, {Fabricant}, {Fazio}, {Brand}, {Brown},
  {Dey}, {Jannuzi}, {Najita}, {McNamara}, {Shields}, \&
  {Rieke}}]{2005ApJS..161....9K}
{Kenter}, A., {Murray}, S.~S., {Forman}, W.~R., {et~al.} 2005, \apjs, 161, 9

\bibitem[{{Kim} {et~al.}(2006){Kim}, {Barkhouse}, {Romero-Colmenero}, {Green},
  {Kim}, {Mossman}, {Schlegel}, {Silverman}, {Aldcroft}, {Anderson}, {Ivezic},
  {Kashyap}, {Tananbaum}, \& {Wilkes}}]{2006ApJ...644..829K}
{Kim}, D.-W., {Barkhouse}, W.~A., {Romero-Colmenero}, E., {et~al.} 2006, \apj,
  644, 829

\bibitem[{{Kim} \& {Fabbiano}(2003)}]{2003ApJ...586..826K}
{Kim}, D.-W. \& {Fabbiano}, G. 2003, \apj, 586, 826

\bibitem[{{La Franca} {et~al.}(2005){La Franca}, {Fiore}, {Comastri}, {Perola},
  {Sacchi}, {Brusa}, {Cocchia}, {Feruglio}, {Matt}, {Vignali}, {Carangelo},
  {Ciliegi}, {Lamastra}, {Maiolino}, {Mignoli}, {Molendi}, \&
  {Puccetti}}]{2005ApJ...635..864L}
{La Franca}, F., {Fiore}, F., {Comastri}, A., {et~al.} 2005, \apj, 635, 864

\bibitem[{{Lehmer} {et~al.}(2005){Lehmer}, {Brandt}, {Alexander}, {Bauer},
  {Schneider}, {Tozzi}, {Bergeron}, {Garmire}, {Giacconi}, {Gilli}, {Hasinger},
  {Hornschemeier}, {Koekemoer}, {Mainieri}, {Miyaji}, {Nonino}, {Rosati},
  {Silverman}, {Szokoly}, \& {Vignali}}]{2005ApJS..161...21L}
{Lehmer}, B.~D., {Brandt}, W.~N., {Alexander}, D.~M., {et~al.} 2005, \apjs,
  161, 21

\bibitem[{{Levenson} {et~al.}(2001){Levenson}, {Weaver}, \&
  {Heckman}}]{2001ApJ...550..230L}
{Levenson}, N.~A., {Weaver}, K.~A., \& {Heckman}, T.~M. 2001, \apj, 550, 230

\bibitem[{{Miyaji} \& {Griffiths}(2002)}]{2002ApJ...564L...5M}
{Miyaji}, T. \& {Griffiths}, R.~E. 2002, \apjl, 564, L5

\bibitem[{{Norman} {et~al.}(2004){Norman}, {Ptak}, {Hornschemeier}, {Hasinger},
  {Bergeron}, {Comastri}, {Giacconi}, {Gilli}, {Glazebrook}, {Heckman},
  {Kewley}, {Ranalli}, {Rosati}, {Szokoly}, {Tozzi}, {Wang}, {Zheng}, \&
  {Zirm}}]{2004ApJ...607..721N}
{Norman}, C., {Ptak}, A., {Hornschemeier}, A., {et~al.} 2004, \apj, 607, 721

\bibitem[{{Page} \& {Carrera}(2000)}]{2000MNRAS.311..433P}
{Page}, M.~J. \& {Carrera}, F.~J. 2000, \mnras, 311, 433

\bibitem[{{Ptak} {et~al.}(2007){Ptak}, {Mobasher}, {Hornschemeier}, {Bauer}, \&
  {Norman}}]{arXiv:0706.1791v1}
{Ptak}, A., {Mobasher}, B., {Hornschemeier}, A., {Bauer}, F., \& {Norman}, C.
  2007, arXiv:0706.1791v1

\bibitem[{{Ranalli} {et~al.}(2003){Ranalli}, {Comastri}, \&
  {Setti}}]{2003A&A...399...39R}
{Ranalli}, P., {Comastri}, A., \& {Setti}, G. 2003, \aap, 399, 39

\bibitem[{{Ranalli} {et~al.}(2005){Ranalli}, {Comastri}, \&
  {Setti}}]{2005A&A...440...23R}
{Ranalli}, P., {Comastri}, A., \& {Setti}, G. 2005, \aap, 440, 23

\bibitem[{{Sarazin} {et~al.}(2000){Sarazin}, {Irwin}, \&
  {Bregman}}]{2000ApJ...544L.101S}
{Sarazin}, C.~L., {Irwin}, J.~A., \& {Bregman}, J.~N. 2000, \apjl, 544, L101

\bibitem[{{Schechter}(1976)}]{1976ApJ...203..297S}
{Schechter}, P. 1976, \apj, 203, 297

\bibitem[{{Schmidt}(1968)}]{1968ApJ...151..393S}
{Schmidt}, M. 1968, \apj, 151, 393

\bibitem[{{Shapley} {et~al.}(2001){Shapley}, {Fabbiano}, \&
  {Eskridge}}]{2001ApJS..137..139S}
{Shapley}, A., {Fabbiano}, G., \& {Eskridge}, P.~B. 2001, \apjs, 137, 139

\bibitem[{{Sullivan} {et~al.}(2004){Sullivan}, {Hopkins}, {Afonso},
  {Georgakakis}, {Chan}, {Cram}, {Mobasher}, \&
  {Almeida}}]{2004ApJS..155....1S}
{Sullivan}, M., {Hopkins}, A.~M., {Afonso}, J., {et~al.} 2004, \apjs, 155, 1

\bibitem[{{Szokoly} {et~al.}(2004){Szokoly}, {Bergeron}, {Hasinger}, {Lehmann},
  {Kewley}, {Mainieri}, {Nonino}, {Rosati}, {Giacconi}, {Gilli}, {Gilmozzi},
  {Norman}, {Romaniello}, {Schreier}, {Tozzi}, {Wang}, {Zheng}, \&
  {Zirm}}]{2004ApJS..155..271S}
{Szokoly}, G.~P., {Bergeron}, J., {Hasinger}, G., {et~al.} 2004, \apjs, 155,
  271

\bibitem[{{Tammann} {et~al.}(1979){Tammann}, {Yahil}, \&
  {Sandage}}]{1979ApJ...234..775T}
{Tammann}, G.~A., {Yahil}, A., \& {Sandage}, A. 1979, \apj, 234, 775

\bibitem[{{Tzanavaris} {et~al.}(2006){Tzanavaris}, {Georgantopoulos}, \&
  {Georgakakis}}]{2006A&A...454..447T}
{Tzanavaris}, P., {Georgantopoulos}, I., \& {Georgakakis}, A. 2006, \aap, 454,
  447

\bibitem[{{Ueda} {et~al.}(2003){Ueda}, {Akiyama}, {Ohta}, \&
  {Miyaji}}]{2003ApJ...598..886U}
{Ueda}, Y., {Akiyama}, M., {Ohta}, K., \& {Miyaji}, T. 2003, \apj, 598, 886

\bibitem[{{White} \& {Ghosh}(1998)}]{1998ApJ...504L..31W}
{White}, N.~E. \& {Ghosh}, P. 1998, \apjl, 504, L31+

\bibitem[{{Wolf} {et~al.}(2004){Wolf}, {Meisenheimer}, {Kleinheinrich},
  {Borch}, {Dye}, {Gray}, {Wisotzki}, {Bell}, {Rix}, {Cimatti}, {Hasinger}, \&
  {Szokoly}}]{2004A&A...421..913W}
{Wolf}, C., {Meisenheimer}, K., {Kleinheinrich}, M., {et~al.} 2004, \aap, 421,
  913

\bibitem[{{Wolf} {et~al.}(2003){Wolf}, {Meisenheimer}, {Rix}, {Borch}, {Dye},
  \& {Kleinheinrich}}]{2003A&A...401...73W}
{Wolf}, C., {Meisenheimer}, K., {Rix}, H.-W., {et~al.} 2003, \aap, 401, 73

\bibitem[{{Zheng} {et~al.}(2004){Zheng}, {Mikles}, {Mainieri}, {Hasinger},
  {Rosati}, {Wolf}, {Norman}, {Szokoly}, {Gilli}, {Tozzi}, {Wang}, {Zirm}, \&
  {Giacconi}}]{2004ApJS..155...73Z}
{Zheng}, W., {Mikles}, V.~J., {Mainieri}, V., {et~al.} 2004, \apjs, 155, 73

\end{thebibliography}
